\documentstyle[aps]{revtex}

\newcommand{\lsi}{\,\raisebox{-0.13cm}{$\stackrel{\textstyle<}
{\textstyle\sim}$}\,}
\newcommand{\gsi}{\,\raisebox{-0.13cm}{$\stackrel{\textstyle>}
{\textstyle\sim}$}\,}
\newcommand{\be}{\begin{equation}}
\newcommand{\ee}{\end{equation}}
\newcommand{\muG}{{\mu {\rm G}}}
\newcommand{\lmpc}{{\lambda_{ \rm Mpc}}}
\newcommand{\lmpcsq}{{\lambda^2_{\rm Mpc}}}
\newcommand{\ud}{\rm Mpc^2/Myr}
\newcommand{\uf}{\rm~ eV^2~ m^{-2}~ s^{-1}~ sr^{-1}}
\begin{document}
\preprint{NYU-TH-00/09/03}
\title{Deducing the Source of Ultrahigh Energy Cosmic Rays}
\author{Glennys R. Farrar}
\address{Department of Physics, New York University, NY, NY 10003,
USA\\ and}
\author{Tsvi Piran}
\address{Racah Institute of Physics, Hebrew University, Jerusalem, Israel
}

\date{\today}
\maketitle
\begin{abstract}
  The apparent lack of suitable astrophysical sources for cosmic rays
  with $E \gsi 10^{19.7}$ eV (UHECRs) is the ``GZK Paradox".  We argue
  that whatever mechanism produces them must also account for events
  down to $\approx 10^{18.7}$ eV, including their isotropy and
  spectral smoothness.  This rules out galactic sources and
  distributed sources such as topological defects; Gamma Ray
  Bursts (GRBs) are unlikely.  We are lead to identify the powerful
  radio galaxy Cen A, at 3.4 Mpc, as the probable source of most
  UHECRs observed at Earth today, and to estimate the extragalactic
  magnetic field to be $\sim 0.3 \muG$.
\end{abstract}
\pacs{
{\tt$\backslash$\string pacs\{\}} }
\narrowtext
\twocolumn
Above an energy of about $10^{19.7}$ eV, cosmic ray protons suffer
inelastic collisions due to photopion production from the cosmic
microwave background\cite{gzk}.  Less than $20\%$ of protons survive
with an energy above $3 \times 10^{20}~(1 \times 10^{20})$ eV for a
distance of 18 (60) Mpc\cite{elbert_sommers}; ultra-high energy (UHE)
nuclei and photons lose energy even more readily.  Yet more than 20
cosmic rays (CRs) have been observed with nominal energies at or above
$10^{20} \pm 30\%$ eV\cite{nw00} 
with the record Fly's Eye event having $3.2 \times 10^{20}$
eV\cite{flyseye}.  There are no apparent suitable astrophysical
sources within the GZK distance to which these events
point\cite{elbert_sommers}, nor is there any significant break in the
spectrum in the GZK region, as would be expected if
higher energy UHECRs are attenuated.  

The CR spectrum can be described by a series of power laws.  At the
``knee", $E \sim 10^{15.5}$ eV, the spectral index steepens from -2.7
to -3.0; at $\sim 10^{17.7}$ eV it steepens further to -3.3. At $E
\sim 10^{18.5}$ eV the spectrum {\it flattens} to an index of
-2.7\cite{yoshida_dai98} and is consistent, within the statistical
uncertainty of the data (which is large above $10^{20}$ eV), with a
simple extrapolation at that slope to the highest energies, possibly
with a hint of a slight accumulation around $10^{19.5}$ eV.  For a
review and references, see \cite{nw00}.

The simplest interpretation of this data is that above $E \sim
10^{18.5}$ eV a new population emerges which dominates the more
steeply falling galactic population, and this new population has an
approximately $E^{-2.7}$ spectrum up to the highest observed energies.
This interpretation is supported by AGASA's analysis of arrival
directions\cite{AGASAang98}: around $10^{18}$ eV the angular
distribution correlates with the galactic center and is consistent
with a galactic origin, while at higher energy this anisotropy
disappears.  The smoothness of the observed spectrum above $10^{18.5}$
eV suggests that a single mechanism is responsible for these events.
Otherwise, there is an apparently miraculous matching of spectra from
different mechanisms, such that the total spectrum is smooth.

We adopt here the following interpretation of the data: {\bf (i)} A
single population of events dominates the CR spectrum from about
$10^{18.5}$ eV to the highest observed energies.  (It will be useful
below to distinguish, within this population, between ``low energy",
$10^{18.7} -10^{19.5}$ eV, and ``high energy", $E\ge 10^{20}$ eV,
UHECRs, avoiding transition regions.) This single mechanism
proposition is extremely powerful.  It means that the high statistics
``low energy'' UHECR data, which have previously been generally
ignored, can be used to draw rather strong conclusions.  {\bf (ii)}
The UHECR population has a smooth spectrum without a substantial
discontinuity at $10^{19.7}$ eV. {\bf (iii)} The UHECRs are nucleons,
although alternative models are mentioned and constrained as well.

Above about $10^{19.7}$ eV, protons display a rapidly falling
attenuation length, plateauing at order 10 Mpc for energies above
$\approx 10^{20.5}$ eV (see e.g. \cite{stanev00}).  Thus independently
of the nature of the sources of UHECRs and of magnetic fields which
may deflect or confine them, ``low energy'' UHECR protons with
pathlengths up to a few Gpc contribute to the flux at Earth, whereas
``high energy'' UHECR protons only reach Earth if their pathlength is
of order few 10's Mpc or less.  We denote the ratio of the average
pathlengths for low and high energy UHECRs the accumulation factor,
$f_{acc}$.  For uniformly distributed sources active over cosmological
times, $f_{acc} \approx $ few Gpc/ few 10's Mpc $\approx 100$.  If the
UHECR sources produce a smooth spectrum, and if Earth is located in a
``typical" environment, the {\it observed spectrum} should have an
offset in normalization between low and high energy UHECRs
approximately equal to the accumulation factor, $f_{acc} \sim 100$.

There seem to be only three ways to account for $f_{acc} \approx 1$ 
in the observed spectrum.
\\
\noindent {\bf (i)} The contribution from sources within
the GZK distance ($\approx 10-20$ Mpc) and corresponding time ($30-60$
Myr) is comparable to the contribution from all other sources active
since redshift of order 1/2 added together. Galactic sources satisfy
this condition trivially, however we will see that isotropy is
difficult to reconcile with a galactic explanation. For extragalactic
sources, this condition implies that our local sources are
significantly more concentrated or more powerful than average.
\\
\noindent  {\bf (ii)} The spectrum at the source could
have the inverse structure to that of the attenuation length.  Some
models appear to give a satisfactory accounting of
the spectrum by this means, but only as a result of requiring a fit
over a too-limited range of energy.  For instance models in which the
spectrum is governed by hadronization of quarks, e.g., the $Z$-burst,
superheavy-relics, and topological defect models, have hard intrinsic
spectra $\sim E^{-1}$.  Thus over the limited energy range $\sim
10^{19.5-20}$ eV, where the attenuation length is rapidly dropping,
the attenuated spectrum can be fit to the observed UHECR spectrum.
However the spectrum of such models has the entirely wrong shape below
the GZK energy, and at $10^{18.7}$ eV, where the flux is well
measured, the model flux is more than an order of magnitude too small.
\\
\noindent {\bf (iii)} GZK energy degradation can be circumvented by
invoking new physics, but models involving GZK-evading
messengers are constrained by this discussion: the nucleon spectrum at
the source must be below that of the messenger particle by a factor
$\approx 1/f_{acc}$, or else the accumulation of those protons
generates an unobserved offset.

A popular possibility considered recently\cite{bbo00,bahcall_waxman}
is a combination of i) and ii), taking advantage of the overdensity of
matter in the local supercluster and hardening the
spectrum\cite{bbo00}.  However the matter overdensity on the GZK scale
falls far short of the $f_{acc}$ level required to remove the offset,
and therefore these efforts fail to account for the spectrum below
$\sim 10^{19.4}$ when the fit at higher energies is acceptable; Fig. 7
of \cite{bbo00} shows that the prediction is more than a factor of 2
too low at $10^{19}$ eV.

In ref. \cite{fp99} we showed that extragalactic magnetic fields can
be of order a few tenths $\muG$ rather than the much smaller values
previously typically assumed in discussions of UHECR propatation, and
we argued that the approximate isotropy of arrival directions could be
due to even the highest energy CRs propagating diffusively.  Now we
explore whether this is consistent with the three conditions outlined
above: {\bf i)} correct directional properties,i.e., sufficient
isotropy {\bf ii)} a spectrum which is approximately a smooth power
law from $\sim 10^{18.7} - 10^{20.5}$ eV, and {\bf iii)} flux of ``low
energy'' UHECRs from the local source should be comparable to that of
all other sources in the Universe.

The nearby radio galaxy Centaurus A (NGC5128) proves to be an
excellent candidate source.  We show that the three conditions exclude
virtually all other alternatives, leading us to conclude that Cen A is
probably the source of most UHE cosmic rays observed on Earth today.
We now summarize some results on diffusion in disordered magnetic
fields and propagation in diffusive media which will be needed below.

Magnetic fields are trapped in ionized matter whose turbulent flow
leads to extensive restructuring of magnetic fields over large and
small distance scales.  It is plausible to assume a Kolmogorov form
for the magnetic field power spectrum as a function of wavenumber,
$B(k) \approx B k^{-11/6}$\cite{bo98,slb98} in this regime; $B$
denotes the magnitude of the field at scale $\lambda$, the maximum
scale of coherent correlations, generally expected to be $ \sim 0.1 -
10$ Mpc.  The Larmor radius of a proton in a magnetic field $B$
orthogonal to its motion is $R_L = 110~ {E_{20}}/{B_{\muG}}$ kpc.

Diffusion has two qualitatively distinct regimes, depending on whether
particles are trapped inside magnetic subdomains or not, causing
different functional dependence on energy of the diffusion coefficient
$D(E)$.  {\it Kolmogorov diffusion} applies when $R_L \ll \lambda$,
i.e., $E< 10^{21} B_\muG\lambda_{Mpc}{\rm eV}$, so the power spectrum
of the magnetic field irregularities is important.  Using the
Kolmogorov spectrum, \cite{bo98} obtained
\be
\label{Dkol} 
D(E) \approx 0.25~
\left( \frac{E_{20}~\lmpcsq }{B_\muG} \right)^{1/3} ~~ {\rm Mpc^2/Myr}.
\ee
As the particle energy increases, there is a transition to {\it Bohm
diffusion} when $R_L \approx \lambda$ and the diffusion
coefficient is of order the Larmor radius times velocity.  In order
that either Bohm or Kolmogorov diffusion apply even for the Fly's Eye
particle, we require $B_\muG \lmpc \gsi 0.3$.  Numerical
simulations\cite{slb98} show that for Bohm diffusion, $D \approx 3~
R_L c$, i.e., $D(E) \approx 0.1~ E_{20}/B_\muG~\ud$.
Since the prefactor in the diffusion coefficient is approximately the
same for either Bohm or Kolmogorov diffusion at the highest energy of
concern here, $E_{20} \approx 3$, we simplify the discussion below to
only Kolmogorov diffusion; given the low statistics above $10^{20}$
eV, this will be adequate.  After infering the pathlength of
propagation, we will check that it is long compared to the
mean-free-path $\approx D/c$, verifying a posteriori the
reasonableness of the diffusive approximation even for the highest
energy CRs.  For the low energy UHECRs, $D(10^{18.7}~{\rm eV}) \approx
1/8~ \ud ~(\lmpcsq 0.3/B_\muG )^{1/3}$.

When $n_0$ CRs are produced by an isotropic source and propagate
diffusively without energy loss, in an environment with a spatially
uniform diffusion coefficient $D$, their number density as a function
of distance $r$ from the source, and time $t$ since emission, is
\be
\label{n(r)}
n(r,t) = \frac{n_0}{(8 \pi~ D~ t)^{3/2}} e^{-r^2/[4 D(E,B) t]}.  
\ee
At a distance $R$ from the source, the number density of CR's of a
given energy increases rapidly with time, reaches a maximum we will
call the diffusion front at $t = R^2/(6 D)$, and then drops slowly
with time $\sim t^{-3/2}$.  Since $D$ is an increasing function of
energy, the diffusion front arrives earlier for higher energies.  In
order to avoid unobserved structure in the energy spectrum, we infer
that the diffusion front of UHECR's from the dominant source must have
already reached Earth, {\it for all energies $10^{18.7 - 20.5}$} eV.  
  
The number of particles with velocity $c$ hitting a unit area in a unit
time in a uniform gas of density $n(r,t)$ is $n(r,t) c$.  Due to the
gradient in the number density with radial distance from the source,
the downward flux at Earth per steradian as a function of angle
$\theta$ to the source is
\be
\label{angdist}
f(\theta,r,t) = (1 + \alpha~ cos ~\theta ) \frac{n(r,t) c}{4 \pi}, 
\ee
where
\be 
\label{alpha}
\alpha = \frac{D(E,B)}{f ~ c}~ \frac{d f}{dr}=\frac{r}{2tc}.
\ee
If the source is collimated into back-to-back pencil beams, eqn
(\ref{n(r)}) is approximately replaced by the superposition of two
sources each with half the strength, separated by a distance $2 d
\approx 2 D/c$ along the direction of collimation and with $t
\rightarrow t - d/c$.  The angular distribution is no longer simply
given by (\ref{angdist}),(\ref{alpha}), but corrections are down by
order $(d/R)^2$.

The result (\ref{alpha}) is very powerful. The anisotropy $\alpha$ is
independent of the diffusion coefficient, as long the approximation of
isotropic source and spatially uniform diffusion coefficient is
adequate.  As a result, {\it the anisotropy is independent of the CR
energy, and the strength and structure of the magnetic field}, so long
as the propagation is diffusive.  This has several important
implications. First, knowledge of $\alpha$ determines the ratio $R/Tc$
where $R$ is the distance between source and Earth and $T$ is the time
between emission and observation at Earth, for instantaneous sources
(see below for extended sources).  Second, UHECRs at all energies can
be combined to improve the statistics in the determination of $\alpha$
without losing its utility for determining $R/Tc $.  The highest
energy events can be excluded, or treated separately, since fields may
not be strong enough that the highest energy events propagate
diffusively.  An energy dependence of the anisotropy signals a
breakdown in this simple description, e.g., due to large scale
structure in the magnetic fields.  The angular distribution is known
to be grossly isotropic, but a quantitative fit to the form
(\ref{angdist}) with respect to an arbitrary direction is not available.

For an AGN, we must integrate eqn (\ref{n(r)}) over the range of
possible propagation times. There is a negligible contribution from
times prior to the arrival time of the diffusion front, so the minimum
propagation time is $t_{min} = {\rm Max}\{ R^2/(3 D),T_{\rm off} \}$,
where $T_{\rm off}$ is the time since the AGN turned off its UHECR
production.  The maximum propagation time is $t_{max} = {\rm
Min}\{\tau_{GZK}), T_{\rm on} \}$, where $\tau_{GZK} \approx 50$ Myr for
$10^{20.5}$ eV and $T_{\rm on}$ is the time since the AGN turned on.
Idealizing the emission to be uniform with a rate $n_0/ \tau$, we
have:
\be
\label{nAGN}
n(r,t) \approx \frac{2~n_0/\tau}{[ 8 \pi D(E,B)]^{3/2}} (t_{min}^{-1/2} -
t_{max}^{-1/2} ).
\ee
Since $D\sim E^{1/3}$ for Kolmogorov diffusion, and shock acceleration
results in a spectrum at the source of $E^{-p}$ with $p$ slightly
greater than 2, the resultant spectral index is close to the $\approx
-2.7$ which is observed.  The data above $E \approx 10^{20}$ eV is
inadequate to decide whether there may be a transition to Bohm
diffusion with its steeper spectrum.  The anisotropy is the
flux-weighted average of $R/2tc$, so $\alpha \approx R/(6 t_{min} c)$,
introducing some energy dependence if $t_{min}$ is determined  by
diffusion rather than the AGN turnoff.

These results allow us to constrain the source of UHECRs observed at Earth.
We must demand that the number density of UHECRs from the proposed local
source be equal to or greater than the accumulated number density from
all UHECR sources in the rest of the Universe:
\be
\label{flux_eq}
n_E \gsi \Gamma_{s} \bar{n} H^{-1} ,
\ee
where $\bar{n}$ is the total number of UHECRs produced by an average
source, $\Gamma_{s}$ is the number of sources per unit volume and unit
time, and $H^{-1} \approx 10^4~ {\rm Myr}$ is the age of the Universe.
For a bursting source such as a GRB, $n_E$ is given by (\ref{n(r)}),
with $n_0 = \kappa \bar{n}$ to allow for an enhancement
factor $\kappa$ if the local source is unusually bright; for an
extended source such as an AGN, $n_E$ is given by (\ref{nAGN}).

We now show that GRBs are unlikely to be responsible for UHECRs, as
follows.  The rate of GRBs is $\Gamma_s = 0.2\times 10^{-3}
~\Gamma_{Sch} /b ~ {\rm Mpc}^{-3}~ {\rm
Myr}^{-1}$\cite{Schmidt99}, where $b \equiv 0.01 b_{0.01} \le 1$ is
the beaming fraction of GRBs;  $b_{0.01}$ may be as small as
$1/2$\cite{Piran99}.  To satisfy 
(\ref{flux_eq}) with (\ref{n(r)}) thus requires $ DT \lsi 0.0012 ~ (
b_{0.01}~\kappa /\Gamma_{Sch})^{2/3} {\rm Mpc}^2$, while $R^2\lsi 6 D
T$ for the diffusion front to have reached Earth, implies $R \lsi 80
~( b_{0.01} ~\kappa /\Gamma_{Sch})^{1/3} {\rm kpc}$.  Thus to dominate
the rest of the GRBs in the Universe, a single GRB would probably have
to be in our galaxy!  But a magnetic field structure capable of
satisfying the isotropy constraint in that case could confine cosmic
rays below the GZK energy from earlier GRBs in the galaxy, and the
accumulation problem is not avoided.  The basic problem is that the
observed rate of GRB's is too high to allow Earth's environment to be
dominated by the output of a local GRB.

AGN's are much rarer than GRBs and the local dominance requirement is
readily satisfied for the values of $D$ and the time scales under
consideration here.  The local flux (\ref{nAGN}) depends on
$t_{min}^{-1/2} - t_{max}^{-1/2}$, which we estimate to be $\sim
0.1-0.2~ {\rm Myr}^{-1/2}$.  The AGN rate is $\approx \rho_{AGN}
\tau_{AGN}/H^{-1}$, where $\rho_{AGN} \approx 10^{-6} {\rm
Mpc^{-3} Myr^{-1}}$ is the number density of AGNs.  (\ref{flux_eq}) is
satisfied without assuming the UHECR output of the local source is
stronger than average.  It is helpful that $\tau_{AGN}$ cancels out,
since it is uncertain.

We now ask for a suitable candidate source.  Defining $D = 0.25~ D_K~
\ud $ and taking $t_{min} < t_{max} \lsi 50 $ Myr, the UHECR diffusion
front is closer than about $\lsi 5 D_K^{1/2}$ Mpc.  It would be very
difficult for the source to be significantly farther than this.  Thus
M87 at 18 Mpc, which has been mentioned as a possible single source of
UHECRs, is too far away if we require diffusive propagation.  Cen A,
at 3.4 Mpc, is by far the nearest powerful radio galaxy; see
ref. \cite{israel} for a comprehensive review.  Already in 1978,
Cavallo\cite{cavallo78} pointed out that the size of its radio lobes
and the strength of its magnetic fields satisfy the Hillas
criterion\cite{hillas} for acceleration of UHECRs.  However at
$b=20^o$ and $l=310^o$, Cen A is in the blind direction of modern
UHECR detectors which are located in the northern hemisphere, and thus
can only be the source if the UHECRs propagate diffusively.  Until
recently there was a prejudice that extragalactic magnetic fields are
of order nG and deflection of UHECRs is negligible, so Cen A was
not considered acceptable.

The observed total luminosity of Cen A is now about $10^{43}$ erg/s,
of which about half is high energy\cite{israel}, so if $\epsilon$ is
the efficiency of UHECR production compared to photon production
extrapolated to $10^{19}$ eV using equal power per decade, we estimate
$E^2 dN/dE/dt \approx 3 \epsilon 10^{53} eV/s$ for $10^{19}$ eV UHECRs.
Using (\ref{nAGN}) and $t_{min}^{-1/2} - t_{max}^{-1/2} = 0.1 {\rm
Myr}^{-1/2}$ to be conservative, we obtain the energy-weighted flux
per str at Earth $E^3 dN/dE \approx 2 \epsilon 10^{25} \uf$, easily consistent
with the observed value of $10^{24.5}\uf$.  Furthermore, the time
scale for evolution of AGNs is $\approx 10$ Myr in general, and of Cen A in
particular\cite{israel}, so that it was likely to have been a powerful
AGN within the most recent GZK time.  If so, its luminosity would have
been $\approx 10^{44} - 10^{45} $ erg/s and the UHECR production
efficiency need only be of order a percent or less.  

The diffusion coefficient for $E = 10^{19}$ eV, $B_\muG = 0.3$ and
$\lmpc = 0.3$, is $0.078 \ud$, so the typical UHECR arrival time from
Cen A is 25 Myr.  The pathlength of a $10^{19}$ eV CR is $\approx 8$
Mpc, many times the diffusion length $D/c \approx 1/4$ Mpc, so the
diffusive approximation applies; the Fly's Eye event barely satisfies
the Bohm diffusion criterion.  The anisotropy $\alpha \approx R/(6
t_{min}c)$ is predicted to be $\approx 0.07$ for $10^{19}$ eV, or less
if the AGN quit emitting UHECRs more than $\approx 25$ Myr ago.

Having identified Cen A to be a good candidate source for UHECRs, we
ask which new physics explanations are consistent with our three
constraints.  Lorentz Invariance violation, such that even $3\times
10^{20}$ eV protons are below threshold for photo-pion production, is
acceptable.  The $Z$-burst model involves messenger neutrinos produced
throughout the Universe, for instance in distant AGNs, but it is
functionally a model with distributed sources, namely the target dark
matter neutrinos, and will be discussed with models of that type
below.  That leaves two GZK-evading messenger models: neutrinos which
interact strongly with nuclei in Earth's atmosphere, or light
gluino-containing baryons whose threshold for photo-pion production is
above $3 \times 10^{20}$ eV.  These models are acceptable only if
nucleon emission by the accelerator is $\lsi 1/f_{acc}$ times the
messenger emission.  If the messenger neutrinos or hadrons are
produced by electromagnetically accelerated protons, as is most
conventional astrophysically, this condition demands considerable
nucleon attenuation in the source.  It is noteworthy that this nucleon
attenuation requirement implies that, if the source were an AGN, it
would necessarily display the spectral characteristics found to be
associated with candidate distant sources in the analysis of
\cite{fb98}.  If the hints of clustering of events and directional
identification with powerful distant matter-enshrouded radio quasars
become real signals with better data, these models would be favored.
If UHECRs are protons from Cen A, these hints are mere fluctuations.

The observed isotropy of ``low" energy UHECRs rules out all models in
which the UHECR sources are proportional to the galactic DM
distribution.  The DM distribution is fairly well known and there is
little flexibility in adjusting its typical clustering scale.  AGASA
compared the anisotropy in the angular distribution of 581 events
above $10^{19}$ eV with that of two popular dark matter distributions
centered on the Milky Way\cite{AGASAang98} and in both cases found
poor agreement with the predicted anisotropy (reduced $\chi^2 \ge
10.0$).  Above $4 \times 10^{19}$ eV the statistics are inadequate to
make this analysis, but with the assumption of a single UHECR
population, that is not necessary to exclude such models.  See
\cite{kuzmin00} for a numerical study which reaches the same
conclusion.

Relaxing the conditions on the source distribution, as may be
appropriate for a new population of astrophysical sources such as
magnetars\cite{bo_magnetars}, topological defects, or eV neutrinos,
does not allow this constraint to be evaded.  To see this, consider a
simple toy model in which all sources are on a shell of radius $d$
around the galactic center, with $d \gg 8.5$ kpc, the distance of
Earth from the galactic center.  This source distribution gives a
dipole anisotropy of $ 0.34 /(d/50{\rm kpc})$.  By increasing $d$ the
anisotropy can be decreased, however large values of $d$ are
unnatural; e.g., an isothermal halo with a core radius of $\sim 1.5$
kpc has $\langle d\rangle \sim 10$ kpc. An upper limit on the
anisotropy implies a lower limit on $d$, but increasing $d$ makes the
local-dominance requirement problematic.

To conclude, we have argued that Earth at this epoch is not exposed to
a generic UHECR spectrum.  The absence of a cutoff at the GZK energy
is a reflection of a coincidental position in space-time relative to
the nearest source.  The most straightforward explanation which
survives the analysis here is that most UHECRs reaching Earth come
from a single, unusually powerful AGN at a distance of a few Mpc.  Cen
A, a powerful radio galaxy at 3.4 Mpc is an excellent candidate.  This
requires diffusion in a magnetic field, probably in the few-tenth
$\mu$b range, which is consistent with observational and theoretical
constraints.  We predict the anisotropy of UHECRs with $E \ge 10^{19}$
eV is of order 7\% or less.  Our analytical discussion should be
followed up by numerical studies, especially to refine our estimates
of the magnetic fields needed to diffuse the highest energy CRs.

The research of GRF was supported in part by NSF-PHY-99-96173.  We
thank T. Kollat, B. Paczynski, M. Rees, R. Sari and S. Yoshida for
valuable comments.


\begin{thebibliography}{10}

\bibitem{gzk}
K.~Greisen.
\newblock {\em Phys. Rev. Lett.}, 16:748, 1966;
G.T. Zatsepin and V.A. Kuzmin.
\newblock {\em Sov. Phys.-JETP Lett.}, 4:78, 1966.

\bibitem{elbert_sommers}
J.~Elbert and P.~Sommers.
\newblock {\em Astrophys. J.}, 441:151, 1995.

\bibitem{nw00}
M.~Nagano and A.~Watson
\newblock {\em Rev. Mod. Phys.}, 72:689, 2000.

\bibitem{flyseye}
D.~J. Bird et~al.
\newblock {\em Astrophys. J.}, 441:144, 1995.

\bibitem{yoshida_dai98}
S.~Yoshida and H.~Dai.
\newblock {\em J. Phys. G: Nucl. Part. Phys.}, 24:905, 1998.

\bibitem{AGASAang98}
N.~Hayashida et~al.
\newblock {\em Astropart. Phys.}, 10:303, 1999.

\bibitem{stanev00}
T.~Stanev et~al.
\newblock astro-ph/0003484.

\bibitem{bbo00}
M.~Blanton, P.~Blasi, and A.~Olinto.
\newblock astro-ph/0009466.

\bibitem{bahcall_waxman}
J.~Bahcall and E.~Waxman.
\newblock astro-ph/9912326.

\bibitem{fp99}
G.~Farrar and T.~Piran.
\newblock {\em Phys. Rev. Lett.}, 84:3527, 2000.

\bibitem{bo98}
P.~Blasi and A.~Olinto.
\newblock {\em Phys. Rev.}, D59:023001, 1999.

\bibitem{slb98}
A.~Lemoine et~al.
\newblock {\em Astropart. Phys.}, 10:141, 1999.

\bibitem{Schmidt99}
M.~Schmidt.
\newblock {\em , Ap. J. in press}.

\bibitem{Piran99}
T.~Piran.
\newblock {\em Physics Reports}, 314:575, 1999.

\bibitem{israel}
F.~P. Israel.
\newblock {\em Astron. Astro. Rev.}, 8:237, 1998.

\bibitem{cavallo78}
G.~Cavallo.
\newblock {\em A \& A}, 65:415, 1978.

\bibitem{hillas}
A.~M. Hillas.
\newblock {\em Ann. Rev. Astron. Astrophys.}, 22:425, 1984.

\bibitem{fb98}
G.~R. Farrar and P.~L. Biermann.
\newblock {\em Phys. Rev. Lett.}, 81:3579, 1998.

\bibitem{kuzmin00}
O.~E. Kalashev et~al.
\newblock astro-ph/0006349.

\bibitem{bo_magnetars}
P.~Blasi and A.~Olinto.
\newblock astro-ph/9912240.

\end{thebibliography}

\end{document}